\documentclass{article}
\usepackage{spconf,amsmath,graphicx}
\usepackage{booktabs}
\usepackage{float}
\usepackage{cite}
\usepackage[T1]{fontenc}
\usepackage{amsfonts}
\usepackage{bm}
\usepackage{algorithm}
\usepackage[noend]{algpseudocode}
\usepackage{amsmath}
\usepackage{amsfonts}
\title{HGCN: harmonic gated compensation network for speech enhancement}
\name{Tianrui Wang$^{\star \dagger}$, Weibin Zhu$^{\star}$, Yingying Gao$^{\dagger}$, Junlan Feng$^{\dagger}$, Shilei Zhang$^{\dagger}$}
\address{$^{\star}$ Institute of Information Science, Beijing Jiaotong University, Beijing, China \\
	$^{\dagger}$ China Mobile Research Institute, Beijing, China}

\begin{document}
	%\ninept
	%
	\maketitle
	\begin{abstract}
		Mask processing in the time-frequency (T-F) domain through the neural network has been one of the mainstreams for single-channel speech enhancement. However, it is hard for most models to handle the situation when harmonics are partially masked by noise. To tackle this challenge, we propose a harmonic gated compensation network (HGCN). We design a high-resolution harmonic integral spectrum to improve the accuracy of harmonic locations prediction. Then we add voice activity detection (VAD) and voiced region detection (VRD) to the convolutional recurrent network (CRN) to filter harmonic locations. Finally, the harmonic gating mechanism is used to guide the compensation model to adjust the coarse results from CRN to obtain the refinedly enhanced results. Our experiments show HGCN achieves substantial gain over a number of advanced approaches in the community.
	\end{abstract}
	\begin{keywords}
	Speech Enhancement, Harmonic, Deep Learning, Pitch
	\end{keywords}
	\section{Introduction}
	\label{sec:intro}
	Speech enhancement aims to improve speech quality by using various algorithms. In recent years, deep learning methods have been applied and achieved promising results in this area. These models could be divided into two main categories, time-domain (T) models and time-frequency domain (T-F) models. T models process the waveform directly to obtain the target speech \cite{Tmodel1}. T-F models precess the spectrum after the short-time fast Fourier transform (STFT) \cite{crn,dccrn,gaze}. Generally speaking, for speech enhancement, it's the T-F structure of speech that is enhanced. In some sense, the processing of the comb harmonic structure of speech constitutes the basis of T-F models \cite{introharmonic1,introharmonic2}. However, in the case of low SNR, the harmonic structure may be masked severely by noise. \cite{phasen} constructs a frequency domain transformation structure to capture harmonic correlations. \cite{introphase2} borrows harmonic enhancement to reconstruct phase. But neither of them explicitly considers the reconstruction of harmonics.

	In principle, the harmonics can be obtained directly from the pitch, while the pitch can be obtained via spectral integral \cite{swipe}. Since the harmonic structure will seldom be completely masked on the spectra even if speech is seriously corrupted by noise, the pitch predicted by spectral integral is reliable. However, \textbf{i)} the frequency resolution after STFT in the deep learning model is fixed and low, which causes the prediction of the pitch by the former spectral integral to be less accurate. And \textbf{ii)} the magnitude values that need to be compensated in the harmonic locations are difficult to be obtained \cite{harmonicmag1,harmonicmag2}.
	
	In this paper, a harmonic gated compensation network (HGCN) is proposed. To tackle challenge \textbf{i)}, we increase the resolution of the pitch candidates and propose a high-resolution harmonic integral spectrum. To tackle challenge \textbf{ii)}, we design a gated \cite{gsr} compensation module to adjust the magnitude of harmonic. In addition, we design a speech energy detector (SED) to do VAD and VRD, which are used to filter harmonic locations. The experimental results show that each sub-module brings a performance improvement, and the proposed method performs better than referenced ones.

	\begin{figure*}[htb]
		\centering
		\vspace{-0.3cm}
		\includegraphics[width=17cm]{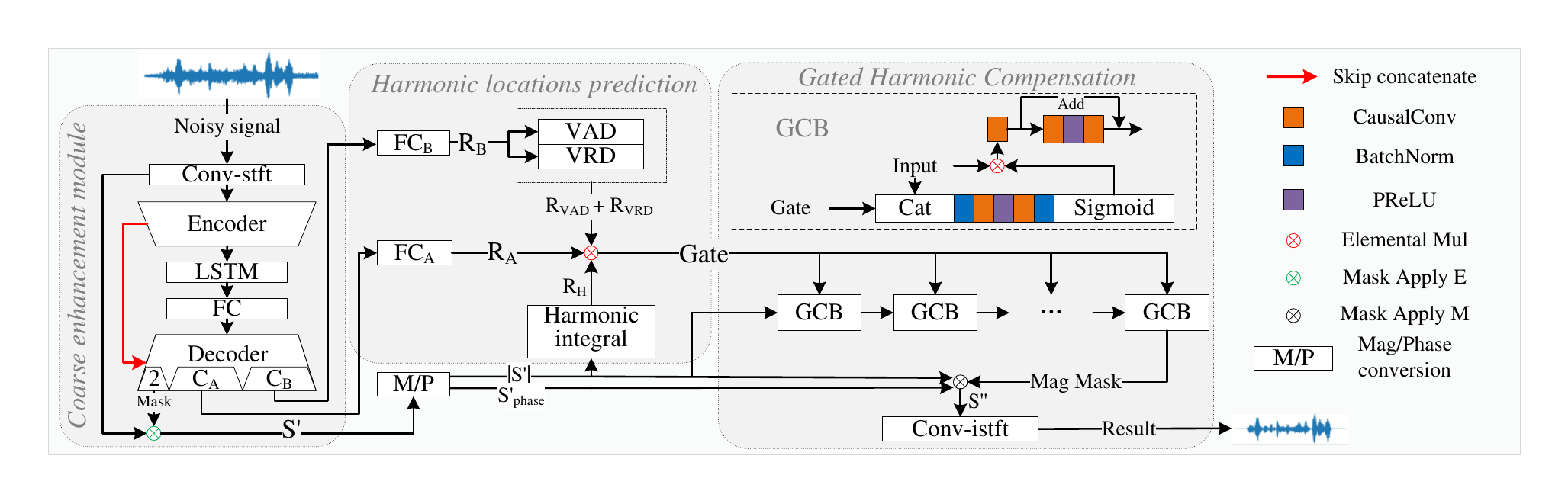}
		\vspace{-0.9cm}
		\caption{Architecture of the proposed HGCN.}
		\vspace{-0.65cm}
		\label{HGCN}
	\end{figure*}
	\section{PROPOSED HGCN}
	\label{sec:model}
	
	The overall diagram of the proposed system is shown in Fig.~\ref{HGCN}. It is mainly comprised of three parts, namely the coarse enhancement module (CEM), harmonic locations prediction module (HM), and gated harmonic compensation module (GHCM). CEM performs a coarse enhancement process on noisy speech. Then HM predicts harmonic locations based on the coarse result of the CEM. GHCM compensates for the coarse result based on the harmonic locations to get the refined result. Each module is described as follows.
	\vspace{-0.01cm}
	
	\subsection{Coarse enhancement module}
	A CRN \cite{crn} model is used to do the coarse enhancement process, which is an encoder-decoder architecture. Specifically, both the encoder and decoder are comprised of Batchnormlization (BN) \cite{batchnorm}, causal 2D convolution blocks (CausalConv) \cite{causalconv}, and PReLU \cite{prelu}. Between the encoder and the decoder, long short-term memory (LSTM) \cite{lstm} is inserted to model the temporal dependencies. Additionally, skip connections are utilized to concatenate the output of each encoder layer to the input of the corresponding decoder layer (red line in Fig.~\ref{HGCN}). Time-domain waveform and T-F spectrum can be interconverted by STFT and inverse transform (iSTFT). In our model, both STFT and iSTFT are implemented by convolution \cite{convstft}. So, the input to the encoder is the noisy complex spectrum, denoted as $\bm{S}=\text{Cat}(\bm{S}_r,\bm{S}_i) \in \mathbb{R}^{T\times 2F}$, where $\bm{S}_r$ and $\bm{S}_i$ represent the real and imaginary parts of the spectrum respectively. And, we compress the input of the encoder with power exponent 0.23 as in \cite{apcsnr}. The decoder predicts a complex ratio mask $\bm{M} = \text{Cat}(\bm{M}_r,\bm{M}_i)\in \mathbb{R}^{T\times 2F}$, where $\bm{M}_r$ and $\bm{M}_i$ represent the real and imaginary parts of mask. We use the mask applying scheme of DCCRN-E \cite{dccrn}, which is called Mask Apply E in Fig.~\ref{HGCN},
	\begin{equation}
		\setlength{\abovedisplayskip}{3pt}
		\setlength{\belowdisplayskip}{2pt}
		\begin{split}
			\bm{S}^{'} &= |\bm{S}| \odot \bm{M}_m \odot e^{j(\bm{S}_{\text{phase}}+\bm{M}_{\text{phase}})} \\
					&= (\bm{S}_r^2+\bm{S}_i^2)^{0.5} \odot \bm{M}_m \odot e^{j\left[\text{arctan}(\bm{S}_i,\bm{S}_r)+\text{arctan}(\bm{M}_i,\bm{M}_r)\right]}
		\end{split}
		\label{maskapplyE}
	\end{equation}
	where $\odot$ denotes the element-wise multiplication operator. $|\cdot|$ and $(\cdot)_{\text{phase}}$ represent the magnitude and phase.  $\bm{M}_m=\text{tanh}\left\{(\bm{M}_r^2+\bm{M}_i^2)^{0.5}\right\}$ is the magnitude mask. $\text{tanh}\left\{\cdot\right\}$ is the activation function proposed in \cite{tanh}. $\text{C}_\text{A}$ and $\text{C}_\text{B}$ in Fig.~\ref{HGCN} are introduced in the next section.

	\subsection{Harmonic locations prediction module}
	\label{sec:harmonic}
	The enhanced result $\bm{S}^{'}$ is first decoupled into $|\bm{S}^{'}|$ and $\bm{S}_{\text{phase}}^{'}$. HM will predict the harmonic locations based on the $|\bm{S}^{'}|$.
	
	There are peaks at integer multiples of the pitch and valleys at half-integer multiples, which are the characteristics of harmonics in the magnitude spectrum. Therefore, the pitch candidates can be set first, and the numerical integral of the multiple positions can be taken as the significance of each candidate. The candidate with the highest significance is the pitch \cite{HPS, swipe}. So, the significance $\bm{Q}$ is calculated as,
	\begin{equation}
		\setlength{\abovedisplayskip}{2pt}
		\setlength{\belowdisplayskip}{2pt}
		\label{si1}
		Q_{t,f}=\sum_{k=1}^{\text{sr}/f}(\frac{1}{\sqrt{k}}\cdot \log{|S^{'}_{t,kf}|}-\frac{1}{\sqrt{k}}\log{|S^{'}_{t,(k-\frac{1}{2})f}}|)
	\end{equation}
	where $\text{sr}$ is half of the audio sample rate. $f$ is the pitch candidate. And $k$ denotes the multiple of the pitch. 
	
\begin{figure}[htb]
		\centering
		\vspace{-0.45cm}
		\includegraphics[width=6cm]{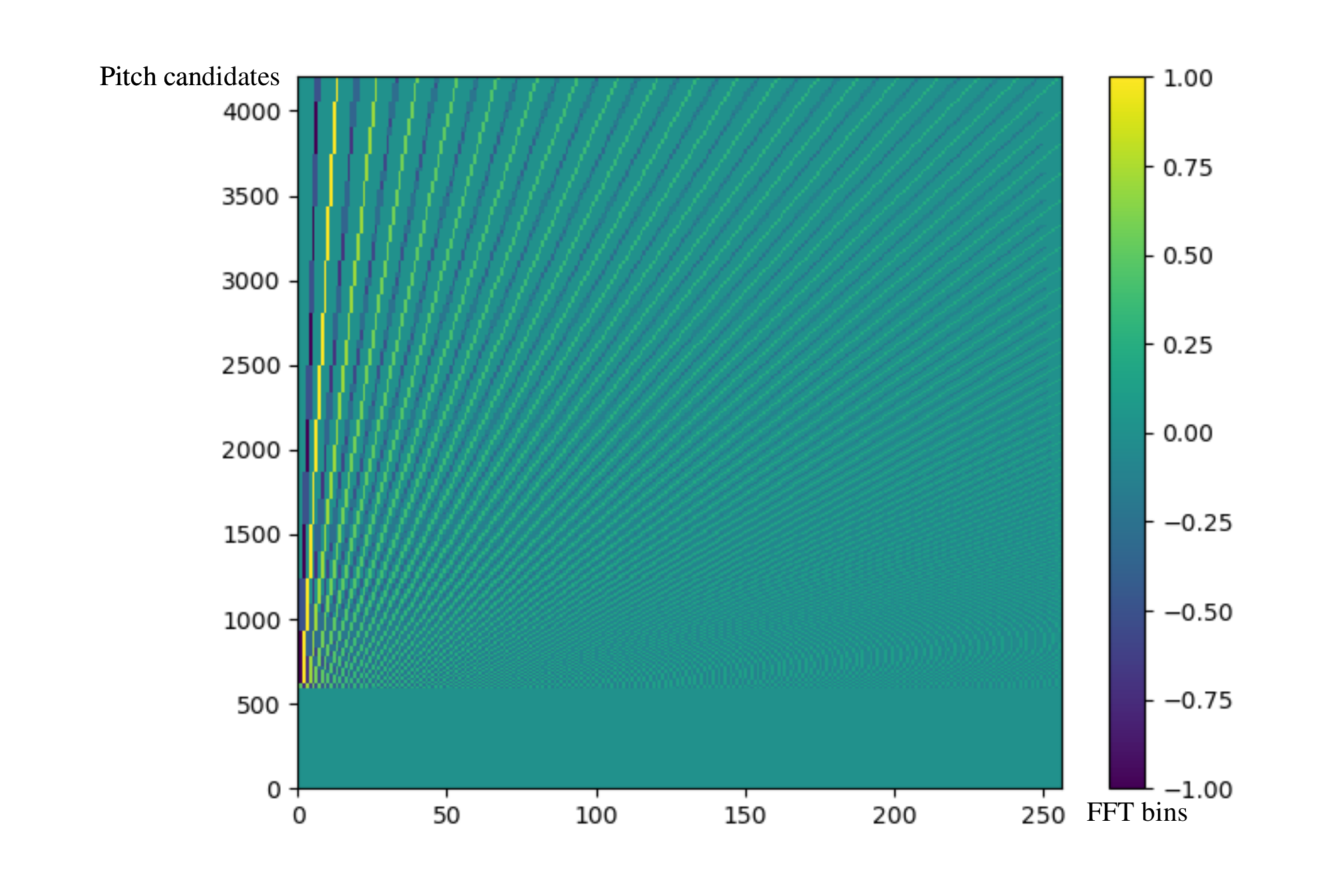}
		\vspace{-0.8cm}
		\caption{High-resolution harmonic integral matrix $\bm{U}$.}
		\vspace{-0.45cm}
		\label{U}
	\end{figure}

	%Since the frequency resolution of the spectral bin is fixed and low, few pitch candidates can be selected.
	For T-F models, 512 Fourier points are often used for audio with 16k sample rate. Since the frequency bandwidth is 31.25~Hz, few pitch candidates can be selected. To solve this problem, a high-resolution integral matrix $\bm{U}$ is designed as Algorithm~\ref{integralspectrum} and Fig.~\ref{U}, where $[\cdot]$ is a rounding operation. We set the pitch candidates with a resolution of 0.1~Hz, and convert the multiple frequencies to the fixed spectral bins. A total of 3600 pitches in 60\textasciitilde 420~Hz (normal pitch range of human) are taken as candidates. Then the Eq.~(\ref{si1}) is improved to
	\begin{equation}
		\setlength{\abovedisplayskip}{2pt}
		\setlength{\belowdisplayskip}{2pt}
		\label{Q}
		\bm{Q}_t = \log{|\bm{S}^{'}_t|} \cdot \bm{U}^\top
	\end{equation}
	where $\bm{Q}_t \in \mathbb{R}^{1\times 4200}$ denotes the pitch candidate significances of the $t$-th frame and the first 600 dimensions are 0. The candidate corresponding to the maximum value in $\bm{Q}_t$ is selected as the pitch, and the corresponding harmonic locations are used as the result $\text{R}_{\text{H}} \in \mathbb{R}^{T\times F}$, where the harmonic locations are 1 and the non-harmonic locations are 0. 
	
\begin{figure*}[htb]
		\centering
		\includegraphics[width=17.4cm]{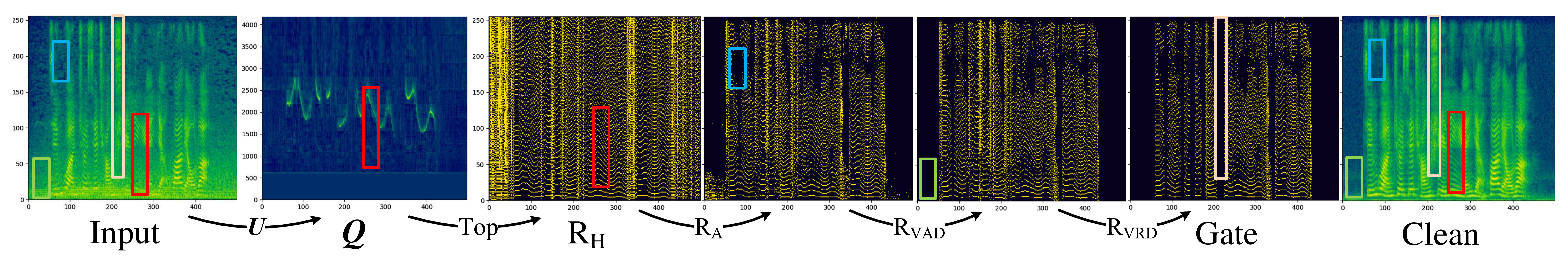}
		\vspace{-0.65cm}
		\caption{The calculation process of harmonic gate.}
		\label{gateprocess}
		\vspace{-0.7cm}
	\end{figure*}

	The pitch and then harmonic locations for each frame can be predicted by Eq.~(\ref{Q}), but in fact, there are no harmonics in non-speech and unvoiced frames, so we apply VAD and VRD to filter $\text{R}_{\text{H}}$ (green and pinkish boxes in Fig.~\ref{gateprocess}). In addition, the energy corresponding to the locations is low even if it's harmonic (blue box in Fig.~\ref{gateprocess}), which need to be filtered out. Therefore, the final harmonic gate is calculated as follows,
	\begin{equation}
		\setlength{\abovedisplayskip}{2pt}
		\setlength{\belowdisplayskip}{3pt}
		\label{gate}
		\text{Gate} =  \text{R}_{\text{VAD}} \odot \text{R}_{\text{VRD}} \odot \text{R}_{\text{A}} \odot \text{R}_{\text{H}}
	\end{equation}
	where $\text{R}_{\text{VAD}} \in \mathbb{R}^{T\times 1}$ and $\text{R}_{\text{VRD}} \in \mathbb{R}^{T\times 1}$ denote the speech activity frames and voiced frames respectively. $\text{R}_{\text{A}} \in \mathbb{R}^{T\times F}$ denotes the non-low energy locations of speech.  $\mathbb{R}^{T\times 1}$ will be copied and expanded into $\mathbb{R}^{T\times F}$.
	% Their calculation process will be described in detail later.

	Both VAD and VRD can be judged based on energy, so we design a speech energy detector to predict two non-low speech energy locations spectra $\text{R}_\text{A}$ and $\text{R}_\text{B}$ with different energy thresholds, where $\bm{\text{R}}_\text{A}$ is designed to filter out the lower energy locations of speech with a smaller threshold, and $\bm{\text{R}}_\text{B}$ is used for VAD and VRD with a larger threshold, which pays more attention to the locations with higher energy. Since the detector needs to be able to resist noise, we change the output channel number of the last CEM decoder to $(2+\text{C}_\text{A}+\text{C}_\text{B})$, where $2$ is the channel number of complex ratio mask for speech enhancement. $\text{C}_\text{A}$ and $\text{C}_\text{B}$ are the channels number of the input $\bm{X}^{'} \in \mathbb{R}^{T\times F \times \text{C}_{\text{A}/\text{B}}}$ for fully connected A ($\text{FC}_\text{A}$) and B ($\text{FC}_\text{B}$) respectively. $\text{FC}_\text{A}$ and $\text{FC}_\text{B}$ output 2-D (low-high) classification probabilities $P_{t,f}=[p_0,p_1]$ for every T-F point $\bm{P} \in \mathbb{R}^{T\times F\times 2}$. And the category can be obtained by $\text{R}_{t,f} = \text{argmax}(P_{t,f})$,	then we can obtain the results of the SED $\text{R}_{\text{A}} \in \mathbb{R}^{T\times F}$ and $\text{R}_{\text{B}} \in \mathbb{R}^{T\times F}$. 
	
\begin{figure}[htb]
		\centering
		\vspace{-0.6cm}
		\includegraphics[width=8.5cm]{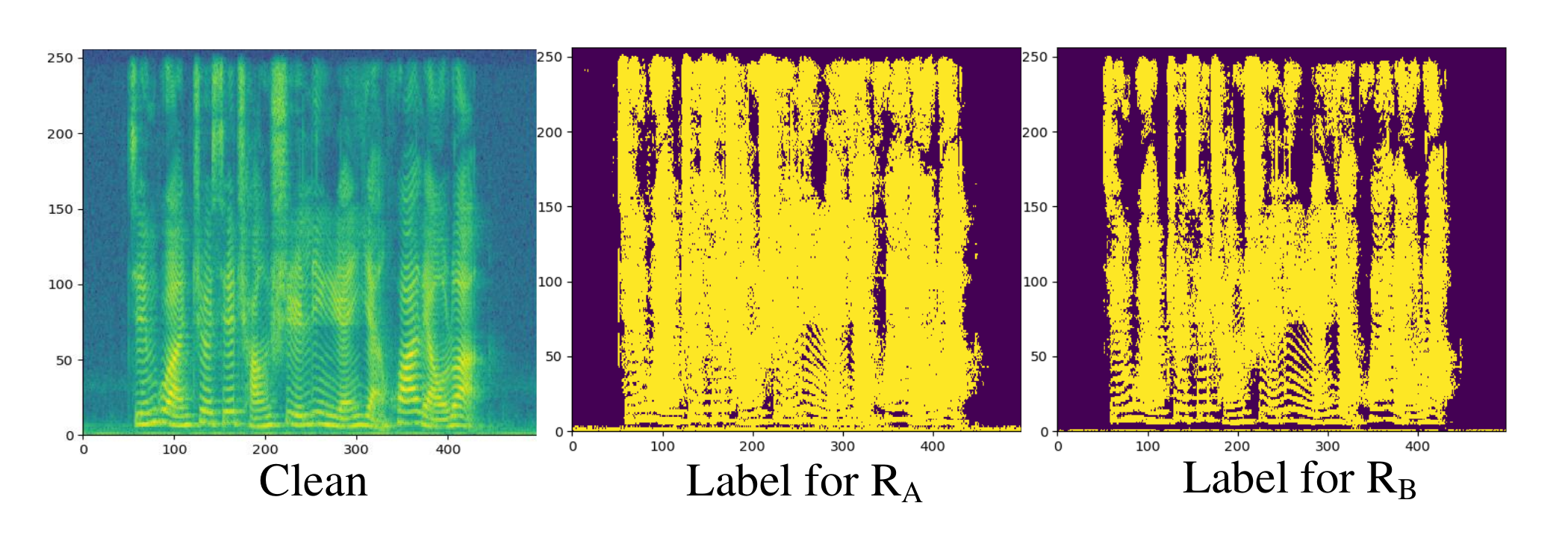}
		\vspace{-0.75cm}
		\caption{Labels for speech energy detection. $\bm{\text{R}}_\text{A}$ is to filter out the lower energy locations with a smaller threshold, $\bm{\text{R}}_\text{B}$ focus on the higher energy locations with a larger threshold.}
		\vspace{-0.7cm}
		\label{labelr}
	\end{figure}

	The labels for the SED are shown in Fig.~\ref{labelr}. We count the mean $\bm{\mu} \in \mathbb{R}^{F \times 1}$ of each bin in the logarithmic magnitude on the clean datas $\dot{\bm{|\mathbb{S}|}}=[|\dot{\bm{S}}|_1,\cdots,|\dot{\bm{S}}|_d] \in \mathbb{R}^{D\times T \times F}$, and standard deviation $\bm{\sigma} \in \mathbb{R}^{Q \times 1}$ of means,
	\begin{equation}
		\setlength{\abovedisplayskip}{3pt}
		\setlength{\belowdisplayskip}{1pt}
		\bm{\mu} = \sum_{i=1}^{D}\left[\left(\sum_{t=1}^{T}\log{|\dot{\bm{\mathbb{S}}}|_{i,t}}\right)/T\right]/D
	\end{equation}
	\begin{equation}
		\setlength{\belowdisplayskip}{1pt}
		\bm{\sigma} = \sqrt{\sum_{i=1}^{D}\left[\left(\sum_{t=1}^{T}\log{|\dot{\bm{\mathbb{S}}}|_{i,t}}\right)/T-\bm{\mu}\right]^2/D}
	\end{equation}
	where $D$ represents the clip number of audio. $|\dot{\bm{S}}|$ represents the magnitude spectra. The energy thresholds $\bm{\kappa}=(\bm{\mu}+\varepsilon \cdot \bm{\sigma})\in \mathbb{R}^{F \times 1}$ of bins are controlled according to different offset values $\varepsilon$ ($\varepsilon_\text{A}=0$ and $\varepsilon_\text{B}=\frac{4}{3}$), and the label for $\bm{\text{R}}_\text{A/B}$ is 1 if the logarithmic magnitude of clean is larger than $\bm{\kappa}$, 0 otherwise.

	Then we can compute $\text{R}_\text{VAD}$ and $\text{R}_\text{VRD}$ based on $\text{R}_\text{B}$,
	\begin{equation}
		\setlength{\abovedisplayskip}{2.5pt}
		\setlength{\belowdisplayskip}{1.5pt}
		(\text{R}_\text{VAD})_{t}=
		\left\{
		\begin{array}{rcl}
			& 1 &, \sum_{f=1}^{F}{(\text{R}_{\text{B}})_{t,f}} > \epsilon  \\
			& 0 &, \sum_{f=1}^{F}{(\text{R}_{\text{B}})_{t,f}} \le \epsilon \\
		\end{array}
		\right.
		\label{VAD}
	\end{equation}
	\begin{equation}
		\setlength{\abovedisplayskip}{1.5pt}
		\setlength{\belowdisplayskip}{1pt}
		\label{voiced}
		(\text{R}_\text{VRD})_{t}=
		\left\{
		\begin{array}{rcl}
			& 0 &, \text{H} > \text{L}  \\
			& 1 &, \text{H} \le \text{L} \\
		\end{array}
		\right.
	\end{equation}
	where $\text{H}=\sum_{f=F/2}^{F}{(\text{R}_{\text{B}})_{t,f}}$ and $\text{L}=\sum_{f=1}^{F/2}{(\text{R}_{\text{B}})_{t,f}}$ denote the number of selected speech points in high and low frequency respectively. $\epsilon$ is the threshold for VAD.
	
	\begin{algorithm}
		\centering
		\caption{Integral matrix}
		\label{integralspectrum}
		\begin{algorithmic}[1]
			\State $\bm{U} \gets \bm{0} \in \mathbb{R}^{4200 \times F}$
			\For{$f \gets 600 \to 4200$}
			\State $\text{last\_index} \gets 0$
			\For{$k\gets 1 \to \left[\text{sr}/(0.1\cdot f)\right]$}
			\State $\text{index}\gets \left[0.1 \cdot f\cdot k\cdot F / \text{sr}\right]$
			\State $\bm{U}_{f,\text{index}}\gets \bm{U}_{f,\text{index}}+(1/\sqrt{k})$
			\If{$\text{index}-\text{last\_index}>1$}
			\State $i \gets [(\text{index}-\text{last\_index})/2]$
			\If{$(\text{index}-\text{last\_index})\mod2\ne 0$}
			\State $\bm{U}_{f,i} \gets \bm{U}_{f,i} -1/(2\sqrt{k})$
			\State $\bm{U}_{f,i+1} \gets \bm{U}_{f,i+1} -1/(2\sqrt{k})$
			\Else 
			\State $\bm{U}_{f,i} \gets \bm{U}_{f,i} -1/\sqrt{k}$
			\EndIf
			\Else 
			\State $\bm{U}_{f,\text{index}} \gets \bm{U}_{f,\text{index}}-1/(2\sqrt{k})$
			\State $\bm{U}_{f,\text{last\_index}}\gets \bm{U}_{f,\text{last\_index}}-1/(2\sqrt{k})$
			\EndIf
			\EndFor
			\EndFor
		\end{algorithmic}
	\end{algorithm}

	\subsection{Gated harmonic compensation module}
	A gated mechanism \cite{gsr} is used to guide the model to compensate for the coarse result $\bm{S}^{'}$ of CEM. The GHCM is composed of multiple gated compensation blocks (GCB) in series to predict the magnitude compensation mask, where GCB is composed of gated convolution layer (GCL) and residual convolution (RC). The input $\bm{X}_{\text{in}}$ of the first GCB is $|\bm{S}^{'}|$, and the subsequent input is the output of the previous one. 
	
	The GCB introduce the gate mechanism during convolution. As shown in Fig.~\ref{HGCN}, we first obtain an attention map $\bm{\alpha} \in \mathbb{R}^{T\times F}$ by concatenating gate Eq.~(\ref{gate}) and the input feature in channel followed by $\text{CB}_{1\times 1}$, 
	\begin{equation}
		\setlength{\abovedisplayskip}{3pt}
		\setlength{\belowdisplayskip}{3pt}
		\bm{\alpha} = \text{sigmoid}\left(\text{CB}_{1\times 1}(\text{Cat}(\text{Gate},\bm{X}_{\text{in}}))\right)
	\end{equation}
	where $\text{CB}_{1\times 1}$ is comprised of BN, CausalConv, and PReLU. Secondly, GCL applies the $\bm{\alpha}$ to the magnitude as $\tilde{\bm{X}} = \bm{X}_{\text{in}} \odot \bm{\alpha}$, then fed $\tilde{\bm{X}}$ into a convolutional layer. Finally, an RC follows the GCL and does a compensation process.
	
	In GCBs, PReLU is used as the activation function, except for the last block which uses sigmoid to predict the compensation magnitude mask $\bm{M}_{\text{GHCM}} \in \mathbb{R}^{T\times F}$. The magnitude mask applying is used and call it Mask Apply M,
	\begin{equation}
		\setlength{\abovedisplayskip}{3pt}
		\setlength{\belowdisplayskip}{2pt}
		\begin{split}
			\bm{S}^{''} &= (|\bm{S}^{'}|+\bm{M}_{\text{GHCM}} \odot |\bm{S}^{'}|) \odot e^{j\bm{S}_{\text{phase}}^{'}} 
		\end{split}
	\end{equation}

	Finally, we convert $\bm{S}^{''}$ into waveform by iSTFT.

	\section{Experiments}
	\subsection{Dataset}
	We evaluate the HGCN on the DNS Challenge (INTERSPEECH 2020) dataset \cite{dns2020}. This dataset includes $500$ hours of clean speech from $2150$ speakers. The noise dataset includes over $180$ hours from $150$ classes. For training, we generate $150$ hours of noisy speech. The SNR is between 0~dB and 40~dB. And data is divided into training and validation set at $4:1$. For testing, the SNR is between $0$~dB and $20$~dB. And the speech data in the testing doesn't participate in the training or validation set, the noises are from \cite{aurora}. A total of $2$ hours of test audio are generated.

	\subsection{Training setup and comparison methods}
	To ensure comparability, we train all models on our dataset with the same setup. The optimizer is Adam \cite{adam}. And the initial learning rate is $0.001$, which will decay $50\%$ when the validation loss plateau for $5$ epochs and the training is stopped if loss plateau for 20 epochs. The kernel size and stride are $(5,2)$ and $(2,1)$. DCRN is utilized as the baseline system. And DCCRN \footnote{https://github.com/huyanxin/DeepComplexCRN} is an improved version of DCRN, which ranked first in the Interspeech2020 DNS challenge real-time-track, so it's utilized as the referenced system.
	
	\textbf{DCRN}: The $32\text{ms}$ Hanning window with $25\%$ overlap and 512-point STFT are used. The channel number of encoder and decoder is $\left\{16,32,64,128,128,128\right\}$. And a 512-units FC layer after a 128-units LSTM is adopted.
	
	\textbf{DCCRN}: The $25\text{ms}$ Hanning window with $25\%$ overlap and 512-point STFT are used. The channel number is $\left\{32,64,128,256,256,256\right\}$, and uses two layers complex LSTM with 128 units for real and imaginary parts respectively. And a dense with 1280 units is after the LSTM. And DCCRN looks ahead one frame in each decoder layer.
	
	\textbf{HGCN(CEM+GHCM+HM)}: The parameter setting of CEM is the same as DCRN, except that the channel number of last decoder is changed to 22 ($\text{C}_\text{A}=\text{C}_\text{B}=10$). Three GCBs are adopted, and their channel numbers and stride are $\left\{8,16,8\right\}$ and $(1,1)$. The $\epsilon$ in Eq.~(\ref{VAD}) is set to 24. We designed the loss functions for $S^{'}$ and $\text{R}_{\text{A}/\text{B}}$ of CEM, $S^{''}$ of GHCM, separately. For $S^{'}$, we use APC-SNR \cite{apcsnr}. For $S^{''}$, we use scale-invariant SNR (SI-SNR) \cite{sisnr} and APC-SNR. For $\text{R}_{\text{A}/\text{B}}$, we use Focal loss \cite{focalloss}. 
	
	\begin{table}[htp]
		\centering
		\vspace{-0.65cm}
		\caption{System comparison on the test set.}
		\begin{tabular}{lcccc}
			\toprule
			\multicolumn{1}{p{4em}}{Model} & \multicolumn{1}{p{2em}}{RTF} & \multicolumn{1}{p{2.5em}}{PESQ} & \multicolumn{1}{p{3em}}{STOI(\%)}  & \multicolumn{1}{p{5em}}{SI-SDR(dB)} \\
			\midrule
			Noisy & - & 1.796 & 93.2 &  10.321 \\	
			DCRN  & \textbf{0.061} & 2.798  & 96.3   &  18.096   \\
			DCCRN & 0.263 & 2.887 & 96.7  & 18.845  \\
			\midrule
			CEM & 0.065 & 2.953 & 96.8 &  18.706  \\
			\ +GHCM & 0.099 & 3.018 & 97.0 &  18.897  \\
			\ \ \ \ \ +HM & 0.109 & \textbf{3.096} & \textbf{97.2} &  \textbf{19.255}  \\
			\bottomrule
			\label{results}
		\end{tabular}
		\vspace{-1.0cm}
		\label{tab:1}
	\end{table}

	\subsection{Experimental results and discussion}
	We compare the performance of HGCN with comparison methods on the test set, and three objective metrics are utilized in the experiments, namely wide band PESQ (PESQ), STOI, and SI-SDR, as shown in Table~\ref{tab:1}. To ensure the generality of the test set. We also did a test on DNS2020 synthetic test set, shown in Table~\ref{tab:dns2020}. Compared with DCRN, the performance of the model has been gradually improved with the gradual addition of the CEM, GHCM, and HM modules. 
	
	The performance of CEM is improved compared to DCRN, which demonstrates the effectiveness of multi-task training \cite{dccrn+}, power compression, and loss function \cite{apcsnr}.
	%	The improvement idea in CEM is similar to that in . Using multi-task training to optimize the model from different perspectives can effectively improve performance. 
	
	GHCM is added on the top of CEM, and only $\bm{\text{R}}_\text{A}$ is used as the gate. Although the performance of the model is improved on all indexes, the improvement ratio of CEM+GHCM on PESQ is greater than that on SI-SDR, even in DNS2020 test set, CEM+GHCM is higher than DCCRN on PESQ, but it is lower on SI-SDR. This is due to that the GHCM compensates for the magnitude and retains the phase of the coarse result. It further causes a slight mismatch between magnitude and phase, while PESQ and STOI only care about the magnitude, SI-SDR will be affected by both magnitude and phase. This is why we add SI-SNR to the loss function of $S^{''}$, otherwise, the effect will be worse. 
	
	HGCN (CEM+GHCM+HM) achieves the best results. We visualize the calculation process of the harmonic gate as shown in Fig.~\ref{gateprocess}. We can observe that the HM can predict the exact harmonic locations, which can better guide the model to compensate for the magnitude spectrum. 
	
	Real Time Factor (RTF) is also tested on a machine with an Intel(R) Core(TM) i5-6200U CPU@2.30 GHz in a single thread (implemented by ONNX). We can observe that the proposed model brings better performance while maintaining good speed.
	
	\begin{table}[htp]
		\centering
		\vspace{-0.65cm}
		\caption{System comparison on DNS-2020 synthetic test set.}
		\begin{tabular}{lccc}
			\toprule
			Model & PESQ & STOI(\%)  & SI-SDR(dB) \\
			\midrule
			Noisy  & 1.582 & 91.5 &  9.071 \\	
			DCRN  & 2.615  & 95.7   &  17.275   \\
			DCCRN & 2.711 & 96.0  & 17.967  \\
			\midrule
			CEM  & 2.753 & 96.1 &  17.539  \\
			\ +GHCM  & 2.812 & 96.3 &  17.841  \\
			\ \ \ \ \ +HM  & \textbf{2.883} & \textbf{96.5} &  \textbf{18.144}  \\
			\bottomrule
			\label{results}
		\end{tabular}
		\vspace{-1.0cm}
		\label{tab:dns2020}
	\end{table}
	
	\section{Conclusion}
	In this paper, to tackle the challenge of speech harmonics being partially masked by noise, a harmonic gated compensation network for monaural speech enhancement is proposed. First, we propose a high-resolution harmonic integral spectrum, which improves the accuracy of harmonic prediction by increasing the resolution of the predicted pitch. In addition, we design VAD and VRD to filter harmonic locations. Finally, the harmonic gating mechanism is used to guide the model to compensate for the coarse results from CRN to obtain the refinedly enhanced result. The experimental results show that the high-resolution harmonic integral spectrum can predict the harmonic locations accurately, and the HGCN performs better than referenced methods.
	
	\bibliographystyle{IEEEbib}
	\bibliography{strings,refs}
	
\end{document}